\documentclass[12pt,english]{article}
\usepackage[T1]{fontenc}
\usepackage[latin9]{inputenc}
\usepackage{geometry}
\usepackage{color}
\usepackage{float}
\usepackage{amsmath}
\usepackage{amsthm}
\usepackage{amssymb}
\usepackage{graphicx}
\usepackage{setspace}
\usepackage[authoryear]{natbib}
\makeatletter

\providecommand{\tabularnewline}{\\}

\theoremstyle{plain}
\newtheorem{prop}{\protect\propositionname}
\theoremstyle{definition}
 \newtheorem{example}{\protect\examplename}
\theoremstyle{plain}
\newtheorem{lem}{\protect\lemmaname}
\theoremstyle{remark}
\newtheorem{rem}{\protect\remarkname}

\allowdisplaybreaks
\usepackage{hyperref}
\hypersetup{
     colorlinks   = true,
     citecolor    = blue
}
\AtBeginDocument{
  
}

\makeatother

\usepackage{babel}
\providecommand{\examplename}{Example}
\providecommand{\lemmaname}{Lemma}
\providecommand{\propositionname}{Proposition}
\providecommand{\remarkname}{Remark}

\begin{document}
\global\long\def\beps{\bar{\varepsilon}}%

\global\long\def\hbeta{\hat{\beta}}%

\title{Can Two Forecasts Have the Same Conditional Expected Accuracy?\thanks{We thank Todd Clark and Mike McCracken for comments on an early version
of the paper.} }
\author{Yinchu Zhu\\
Department of Economics,\\
Brandeis University\and Allan Timmermann\\
Rady School of Management,\\
University of California, San Diego \\
}
\maketitle
\begin{abstract}
The approach for testing equal predictive accuracy for pairs of forecasting
models proposed by \citet{giacomini2006tests} assumes that the parameters
of the underlying forecasting models are estimated using a rolling
window of fixed width and incorporates the effect of parameter estimation
in the null hypothesis. We show that a necessary and sufficient condition
for the conditionally expected loss differential of two forecasting
models to be a martingale difference sequence is that the outcome
is a simple average of the two forecasts. When the forecasts contain
parameter estimation errors, this means that the conditional mean
of the outcome has to be a function of past estimation errors--a
condition that fails in many situations. We also show that the null
can fail even in the absence of parameter estimation for many types
of stochastic processes in common use.
\end{abstract}

\section{Introduction}

In an important contribution to the literature on economic forecasting,
\citet{giacomini2006tests} (GW, henceforth) develop a novel approach
for comparing the accuracy of alternative economic forecasts and test
the null of equal expected predictive accuracy. GW incorporate the
effect of parameter estimation in the null hypothesis and assume that
estimation error does not vanish asymptotically by requiring that
fixed-length windows are used to estimate the parameters of the underlying
forecasting models. Using this approach, simple tests of equal predictive
accuracy do not have a degenerate limiting distribution even in comparisons
of nested forecasting models, thus addressing a key problem causing
difficulties for earlier tests.\footnote{See, e.g., \citet{west1996asymptotic}, \citet{clark2001tests}, \citet{Clark2005a}
and \citet{mccracken2007asymptotics}.} 

The GW test has found widespread use in applied work in economics
and has become the standard method for comparing the predictive accuracy
of nested forecasting models while accounting for the effect of parameter
estimation.\footnote{As of end-May, 2020, \citet{giacomini2006tests} has nearly 1,400
Google Scholar citations.} It is, therefore, important to establish conditions under which the
null hypothesis entertained by GW is valid and so can be used to compare
the predictive accuracy of alternative forecasts. The null hypothesis
in GW is that the expected loss differential, i.e., the difference
between the expected loss of a pair of forecasts, is a martingale
difference sequence (MDS) conditional on some information set which
typically includes, at a minimum, current and past observations of
the outcome and data used to generate the forecasts.

We show here that the MDS property conditional on past data fails
to hold in many situations when the forecasts are generated using
a set of estimated model parameters as assumed in the analysis of
GW. This conclusion holds regardless of whether a fixed-width rolling
window or an expanding window is used to estimate the parameters of
the underlying forecasting models and regardless of whether the models
are nested or non-nested. Caution should therefore be exercised when
applying the conditional GW test to compare the performance of forecasting
models. Even in the absence of parameter estimation error, we show
that the MDS property fails whenever the underlying data generating
process (DGP) for the outcome does not satisfy a restrictive finite
dependence condition that rules out many standard models used in applied
work.

This conclusion has important practical implications. In particular,
there can be substantial autocorrelation in the loss difference and
we show that the GW test can result in severe size distortions for
testing an unconditional null using popular schemes for estimating
the long-run variance of the loss differential. We also establish
a valid procedure for testing that the null of equal predictive accuracy
of two rolling-window forecasts holds ``on average''. This sub-sampling
t-test approach uses a self-normalization structure and avoids directly
constructing robust standard errors that account for serial correlation
in the loss differentials.

It is worth emphasizing that our analysis pertains to tests of the
null hypothesis that the expected loss differential is mean-zero conditional
on the parameter estimates used by the underlying forecasting models
and possibly other information. It does not apply to tests of the
null that the unconditional mean of the loss differential equals zero.
This unconditional null does not require that two forecasts are equally
accurate at each point in time but only that they are equally accurate
on average. Testing this unconditional null is more common in practice
and is typically conducted using the popular Diebold-Mariano test.

The outline of our analysis is as follows. Section 2 introduces our
setup and demonstrates that the GW null fails to be valid in the presence
of estimated parameters and also fails in the absence of a restrictive
finite dependence condition on the DGP. Second 3 pursues practical
implications of our theoretical results, while Second 4 analyzes tests
of the null that the forecasts have the same unconditionally expected
loss, and Section 5 concludes. Proofs are contained in an appendix.

\section{The null of equal conditional expected predictive accuracy}

This section introduces the forecast environment and demonstrates
that the GW null is not, in general, appropriate for comparing the
conditionally expected loss of a pair of forecasting models.

\subsection{Setup}

Our forecast environment closely mirrors the setup in GW. We are interested
in comparing the predictive accuracy of a pair of one-step-ahead forecasts
$f_{1,t},f_{2,t}$ of some variable $Y_{t+1}$.\footnote{For simplicity, we restrict the forecast horizon to a single period,
but our results are easily generalized to arbitrary horizons of finite
length.} Each forecast is generated using information available at time $t\mbox{,}$
$\mathcal{F}_{t}=\sigma(W_{1}^{\prime},W_{2}^{\prime},...,W_{t}^{\prime})$,
where $W_{t}=(Y_{t},X_{t}^{\prime})^{\prime}$, and $X_{t}$ is a
set of predictor variables. Hence, $\mathcal{F}_{t}$ contains current
and past values of the outcome, forecast and predictors. GW carefully
state that the forecasts are adapted to the most recent $m$ values
of $W_{t}$, i.e., $f_{i,t}=f(W_{t},W_{t-1},...,W_{t-m+1};\hat{\beta}_{i,t,m})$.\footnote{Note that the estimation window, $m$, can differ for the two forecasting
models, i.e., we can have $m_{1},m_{2}$ and define $m=max(m_{1},m_{2})$
and all results continue to go through. } The setup includes, but is not limited to, linear prediction models
of the form $y_{t+1}=\beta_{i}^{\prime}X_{it}+\varepsilon_{it+1},$estimated
by least squares, i.e., $\hbeta_{i,t,m}=\left(\sum_{s=t-m}^{t-1}X_{is}X_{is}^{\prime}\right)^{-1}\left(\sum_{s=t-m}^{t-1}X_{is}y_{s+1}\right)$.

The precision of the forecasts is evaluated using a loss function,
$L(y_{t+1},f_{i,t})$ which is a mapping from the space of outcomes
and forecasts to the real line. Under the commonly used squared error
loss, $L(y_{t+1},f_{i,t})=e_{i,t+1}^{2}$, where $e_{i,t+1}=y_{t+1}-f_{i,t}$
($i=1,2$) is the forecast error. Furthermore, following \citet{Diebold1995},
define the loss differential as the loss of model 1 relative to that
of model 2: 
\begin{equation}
\Delta L_{t+1}\equiv\Delta L_{t+1}(y_{t+1},f_{1,t},f_{2t})=L(y_{t+1},f_{1,t})-L(y_{t+1},f_{2,t,m}).\label{eq:delta L}
\end{equation}

Under squared error loss, 
\begin{equation}
\Delta L_{t+1}=(y_{t+1}-f_{1,t})^{2}-(y_{t+1}-f_{2,t})^{2}.\label{eq:MSE loss}
\end{equation}

The null hypothesis considered by GW is that, conditional on some
information set, $\mathcal{G}_{t}$, the loss differential is a martingale
difference sequence (MDS): 
\begin{equation}
E[\Delta L_{t+1}(y_{t+1},f_{1,t},f_{2t})|\mathcal{G}_{t}]=0,\label{eq:MDS null}
\end{equation}
almost surely for $t=1,2,...$. GW write that ``Note that we do not
require $\mathcal{G}_{t}=\mathcal{F}_{t}$, although this is a leading
case of interest...'' For simplicity, in the rest of the paper, we
focus on this leading case with $\mathcal{G}_{t}=\mathcal{F}_{t}$.

Forecast comparisons are often conducted using (pseudo) out-of-sample
evaluation methods which split a sample of $T$ observations into
an initial sample using $m$ observations for parameter estimation
and a forecast evaluation sample which consists of the remaining $n$
observations, so $T=n+m$.\footnote{Out-of-sample forecast evaluations can have substantially weaker power
than full-sample tests (see, e.g., \citet{Inoue2005} and \citet{Hansen2015}),
but are less prone to data mining biases (\citet{Hansen2015b}) and
can provide important information about the time-series evolution
in a prediction model's performance and its value in real time.} GW assume that $m$ is bounded by a finite constant although, in
general, we can regard $m=m_{T}$ and $n=n_{T}$ as functions of $T$.
For notational simplicity, we suppress the subscripts.

To test the null in (\ref{eq:MDS null}), GW propose the test statistic\footnote{For simplicity, we choose the instrument to be a constant. }
\begin{equation}
GW_{T}=\frac{\sum_{t=m}^{T-1}\Delta L_{t+1}}{\sqrt{\sum_{t=m}^{T-1}(\Delta L_{t+1})^{2}}}.\label{eq: GW stat}
\end{equation}
The simple expression in the denominator exploits the property that,
under the null (\ref{eq:MDS null}) that $\Delta L_{t+1}$ follows
a MDS, there is no need to correct for possible serial correlation
in $\Delta L_{t+1}$. 

To test the hypothesis that $E[\Delta L_{t+1}\mid\mathcal{F}_{t}]=0$
almost surely, GW point out that one can choose any random variable
$h_{t}\in\mathcal{F}_{t}$ and test the unconditional moment condition
$E(\Delta L_{t+1}h_{t})=0$. However, when $P(E[\Delta L_{t+1}\mid\mathcal{F}_{t}]=0)<1$,
there exists a variable $h_{t}^{*}\in\mathcal{F}_{t}$ such that $E(\Delta L_{t+1}h_{t}^{*})\neq0$.
Hence, results in GW hinge on the MDS condition which we examine in
this paper. 

\subsection{Implications of the MDS null}

We first establish a necessary and sufficient condition for the validity
of the null hypothesis in (\ref{eq:MDS null}): 
\begin{prop}
\label{prop: master prop}Assume that $f_{1,t}\neq f_{2,t}$ almost
surely. Then under squared error loss $E(\Delta L_{t+1}\mid\mathcal{F}_{t})=0$
almost surely if and only if $E(y_{t+1}\mid\mathcal{F}_{t})=(f_{1,t}+f_{2,t})/2$
almost surely.
\end{prop}
This result shows that the conditionally expected loss differential
follows an MDS process if and only if the conditional expectation
of the outcome is a simple average of the two forecasts. When the
forecasts contain parameter estimation errors, this means that the
conditional mean of $y_{t+1}$ has to be a function of past estimation
errors. This is quite unnatural as DGPs are typically considered to
be objective processes whose dynamics is not related to an estimation
scheme.

Proposition \ref{prop: master prop} can be used constructively to
establish cases in which the null hypothesis in (\ref{eq:MDS null})
is valid, as we next show. 
\begin{example}
Suppose that \textit{$y_{t+1}=\rho y_{t}+\varepsilon_{t+1}$ with
$E[\varepsilon_{t+1}|\mathcal{F}_{t}]$=0, so $E[y_{t+1}\mid\mathcal{F}_{t}]=\rho y_{t}$.
Moreover, if $f_{1,t}=(\rho-\delta)y_{t}$ and $f_{2,t}=(\rho+\delta)y_{t}$,
then $E[y_{t+1}|\mathcal{F}_{t}]=\frac{1}{2}(f_{1,t}+f_{2,t})$ so
that, from Proposition }\ref{prop: master prop},\textit{ the squared
error loss differential follows an MDS process.} 
\end{example}
A popular practice is to estimate the parameters of the forecasting
model using the most recent $m$ observations. In our leading case
as well as in GW, $m$ is fixed or bounded, so that estimation errors
do not vanish, while the asymptotic analysis lets $n\rightarrow\infty$.
As we shall see, this case can never satisfy $E(y_{t+1}\mid\mathcal{F}_{t})=(f_{1,t}+f_{2,t})/2$
almost surely without a restrictive finite dependence condition and
so the MDS condition (\ref{eq:MDS null}) cannot hold.
\begin{example}
As a simple example with estimation error, let $\bar{x}_{i,t}=m^{-1}\sum_{s=0}^{m-1}x_{i,t-m}$
for $i=1,2$ be $m$-period moving averages, while 
\[
y_{t+1}=\frac{1}{2}\bar{x}_{1,t}+\frac{1}{2}\bar{x}_{2,t}+\varepsilon_{t+1},
\]
where $\varepsilon_{t+1}$ is an MDS so $E[\varepsilon_{t+1}|\mathcal{F}_{t}]=0$.
Setting $f_{1,t}=\bar{x}_{1,t}$ and $f_{2,t}=\bar{x}_{2,t}$, both
forecasts use misspecified (under-dimensioned) forecasting models
that omit one moving average component. However, once again we have
$E[y_{t+1}|\mathcal{F}_{t}]=\frac{1}{2}\bar{x}_{1,t}+\frac{1}{2}\bar{x}_{2,t}=\frac{1}{2}(f_{1,t}+f_{2,t})$,
so it follows from Proposition \ref{prop: master prop} that the squared
error loss follows an MDS process. \footnote{We thank a referee for suggestion these two examples.}
\end{example}
Even for forecasts that do not depend on estimated parameters, we
can show that the MDS property fails whenever the underlying DGP does
not satisfy the restrictive finite dependence condition. In particular,
let $\mathcal{F}_{t-m:t}$ denote the $\sigma$-algebra generated
by data from time $t-m$ to $t$. All forecasts based on a rolling
window of size $m$ are $\mathcal{F}_{t-m:t}$-measurable; this allows
for any estimation methodology ranging from the basic least-squares
estimators to sophisticated machine learning methods. Recall that
$\mathcal{F}_{t}$ denotes the $\sigma$-algebra generated by all
the data up to time $t$.

We next show that the MDS condition (\ref{eq:MDS null}) cannot hold
for $\Delta L_{t+1}$ if the outcome variable $y_{t+1}$ does not
have finite dependence:\textbf{ }
\begin{prop}
\label{prop: impossibility rolling window}Assume that $f_{1,t}\neq f_{2,t}$
almost surely and $P\left(E(y_{t+1}\mid\mathcal{F}_{t})=E(y_{t+1}\mid\mathcal{F}_{t-m:t})\right)<1$.
If $f_{1,t}$ and $f_{2,t}$ are both $\mathcal{F}_{t-m,t}$-measurable,
then $\{\Delta L_{t+1},\mathcal{F}_{t}\}$ is not a martingale difference
sequence, i.e., $P(E(\Delta L_{t+1}\mid\mathcal{F}_{t})=0)<1$.
\end{prop}
The finite-dependence condition in Proposition \ref{prop: impossibility rolling window}
requires that the conditional mean of $y_{t+1}$ only depends on the
most recent $m$ data points. This condition is ruled out by many
widely used models that are not finite-order Markov such as MA and
ARMA processes as well as unobserved components (state space) models
and GARCH-in-mean processes.

\subsection{MDS Property and Estimation Scheme}

Proposition \ref{prop: impossibility rolling window} shows that unless
$E(y_{t+1}\mid\mathcal{F}_{t})=E(y_{t+1}\mid\mathcal{F}_{t-m:t})$
with probability one, the loss difference cannot be an MDS, no matter
how one computes the forecasts with or without estimated parameters
and regardless of the estimation method. The result also does not
require us to make specific assumptions on the DGP such as stationarity.
Hence, in the absence of the restrictive finite dependence condition,
there is an inherent contradiction between the MDS property and rolling
window estimation. 

In most forecasting problems, the main challenge is to model and estimate
the conditional mean $E(y_{t+1}\mid\mathcal{F}_{t})$. However, Proposition
\ref{prop: master prop} says that the MDS null is equivalent to $E(y_{t+1}\mid\mathcal{F}_{t})=(f_{1,t}+f_{2,t})/2$.
Since $f_{1,t}$ and $f_{2,t}$ are computed from a fixed-width rolling
window, this result is saying that the true conditional mean $E(y_{t+1}\mid\mathcal{F}_{t})$
can be learned from the data without letting the sample size increase. 

This is quite a restrictive setup. Even if we have a correctly specified
parametric model $f(W_{1:t};\theta)$ for the entire distribution
$y_{t+1}\mid\mathcal{F}_{t}$ with parameter $\theta$, the optimal
estimator (i.e., the maximum likelihood estimate) for $\theta$ in
the rolling window of length $R$ would have an error of rate $O(R^{-1/2})$.
Hence, having no estimation error at all for $E(y_{t+1}\mid\mathcal{F}_{t})$
for fixed $R$ is hard to imagine. 

In the following AR$(q)$ example, we show that the MDS null fails
to hold under realistic estimation schemes with finite estimation
windows.
\begin{example}
\label{exa: AR}$Y_{t+1}=\sum_{j=1}^{q}\rho_{j}Y_{t+1-j}+\varepsilon_{t+1}$,
where $\varepsilon_{t+1}$ is i.i.d with $E(\varepsilon_{t+1})=0$.
Thus, the $\sigma$-algebra from the observed data is $\mathcal{F}_{t}=\sigma(Y_{t},Y_{t-1},...)$.
Suppose that $f_{1,t}=X_{t,1}^{\prime}\hbeta_{t,1}$ and $f_{2,t}=X_{t,2}^{\prime}\hbeta_{t,2}$,
where $X_{t,1}\in\mathbb{R}^{q_{1}}$ and $X_{t,2}\in\mathbb{R}^{q_{2}}$
with $\max\{q_{1},q_{2}\}\leq q$ are vectors with elements in $Y_{t-q:t}:=(Y_{t},...Y_{t-q})'$.
For $j\in\{1,2\}$, we assume that $\hbeta_{t,j}=g_{j}(Y_{t},Y_{t-1},...,Y_{t-m_{j}})$
for some measurable function $g_{j}$, where $m_{j}>q_{j}$ is finite.
Notice that some components of $\hbeta_{t,j}$ are allowed to be zero;
hence, without loss of generality, we can assume that $X_{t,1}=X_{t,2}=Y_{t-q:t}$. 
\end{example}
Building on this example, we have the following result:
\begin{lem}
\label{lem: AR example}Suppose that $Y_{t+1}=\sum_{j=1}^{q}\rho_{j}Y_{t+1-j}+\varepsilon_{t+1}$,
where $\varepsilon_{t+1}$ is i.i.d with $E(\varepsilon_{t+1})=0$.
Assume that $f_{1,t}\neq f_{2,t}$ almost surely and that $\varepsilon_{t+1}$
has a density. If $P\left({\rm Var}(\hat{\beta}_{t,1}+\hbeta_{t,2}\mid Y_{t-q:t})>0\right)>0$,
then $\{\Delta L_{t+1},\mathcal{F}_{t}\}$ is not a martingale difference
sequence. 
\end{lem}
The requirement that $P\left({\rm Var}(\hat{\beta}_{t,1}+\hbeta_{t,2}\mid Y_{t-q:t})>0\right)>0$
almost surely is very weak and only rules out cases in which ${\rm Var}(\hat{\beta}_{t,1}+\hbeta_{t,2}\mid Y_{t-q:t})$
is never positive, for sure. Typically, the estimation window is larger
than the number of lags $q$. Therefore, we should expect that entries
$Y_{s}$ for $s<t-q$ will affect $\hat{\beta}_{t,1}+\hbeta_{t,2}$.
As a result, conditional on $Y_{t-q:t}$, we should expect non-zero
variations in $\hat{\beta}_{t,1}+\hbeta_{t,2}$. By Lemma \ref{lem: AR example},
for autoregressive DGPs with $q$ lags, we should not expect the MDS
null to hold if the estimation window is larger than $q$ under commonly
used estimators (OLS, MLE or GMM). 

\subsection{Nested models}

For the nested case, consider the DGP 
\begin{equation}
y_{t+1}=\beta'X_{t}+\varepsilon_{t+1},\label{eq:nested DGP}
\end{equation}
where $X_{t}=(1,x_{t})'\in\mathbb{R}^{2}$ is a vector of fixed regressors
and $\varepsilon_{t+1}$ is i.i.d from $N(0,\sigma^{2})$. Consider
the forecasts from a big model (intercept and $x_{t}$) and a small
model (intercept only): $f_{1,t}=X_{t}'\hbeta_{t,m}$ with $\hbeta_{t,m}=\left(\sum_{s=t-m}^{t-1}X_{s}X_{s}^{\prime}\right)^{-1}\left(\sum_{s=t-m}^{t-1}X_{s}y_{s+1}\right)$
and $f_{2,t}=m^{-1}\sum_{s=t-m}^{t-1}y_{s+1}$. 
\begin{prop}
\label{lem: size DGP}Let $X_{t}=(1,x_{t})'\in\mathbb{R}^{2}$, $X_{(t-1)}=(X_{0},\cdots,X_{t-1})'\in\mathbb{R}^{t\times2}$,
$x_{(t-1)}=(x_{0},\cdots,x_{t-1})'\in\mathbb{R}^{t}$ and $1_{t}=(1,\cdots,1)'\in\mathbb{R}^{t}$.
Suppose that
\begin{equation}
c^{2}=\sigma^{2}\left[\sum_{t=m+1}^{m+n}\left(X_{t-1}^{\prime}\left(X_{(t-1)}^{\prime}X_{(t-1)}\right)^{-1}X_{t-1}-\frac{x_{t}^{2}}{x_{(t-1)}'x_{(t-1)}}\right)\right]/\left[\sum_{t=m+1}^{m+n}\left(1-\frac{x_{(t-1)}'1_{t-1}}{x_{(t-1)}'x_{(t-1)}}x_{t}\right)^{2}\right].\label{eq:lemma3}
\end{equation}
Then, under squared error loss (\ref{eq:MSE loss}),
\begin{enumerate}
\item $E\left[\sum_{t=m+1}^{m+n}\Delta L_{t}\right]=0$. 
\item $\{\Delta L_{t+1},\mathcal{F}_{t}\}$ is not a Martingale Difference
Sequence, i.e., $P\left[E(\Delta L_{t+1}\mid\mathcal{F}_{t})=0\right]<1$. 
\end{enumerate}
\end{prop}
Proposition \ref{lem: size DGP} holds for any sample size and has
two implications. First, the MDS condition fails when $m$ is fixed
(rolling estimation window) and the length of the out-of-sample period
($n$) tends to infinity. Second, the MDS condition also fails when
both $m$ and $n$ tend to infinity, i.e., with an expanding estimation
window.

\subsection{Non-nested models}

For the case with non-nested models consider the DGP 
\begin{equation}
y_{t+1}=\beta_{1}x_{1t}+\beta_{2}x_{2t}+\varepsilon_{t+1},\label{eq:DGP non-nested}
\end{equation}
where $E(\varepsilon_{t+1}\mid\mathcal{F}_{t})=0$. We use $f_{1,t}=\hbeta_{1t}x_{1t}$
and $f_{2,t}=\hbeta_{2t}x_{2t}$, where $\hbeta_{it}=\left(\sum_{s=t-m}^{t-1}x_{is}^{2}\right)^{-1}\left(\sum_{s=t-m}^{t-1}x_{is}y_{s+1}\right)$
for $i\in\{1,2\}$. The following result states that the MDS condition
(\ref{eq:MDS null}) fails:
\begin{prop}
\label{prop: non-nested model}Consider the DGP in (\ref{eq:DGP non-nested})
and assume squared error loss (\ref{eq:MSE loss}). Then (1) If $P(\beta_{1}x_{1t}\neq\beta_{2}x_{2t})=1$
and $P(\beta_{1}x_{1t}+\beta_{2}x_{2t}=0)\neq1$, we do not have $P\left[E(\Delta L_{t+1}\mid\mathcal{F}_{t})=0\right]=1$.\\
(2) Let $f_{1,t}=\hbeta_{1t}x_{1t}$ and $f_{2,t}=\hbeta_{2t}x_{2t}$
with estimates $\hbeta_{1t}$ and $\hbeta_{2t}$ being $\mathcal{F}_{t}$-measurable.
If $P(\hbeta_{1t}x_{1t}\neq\hbeta_{2t}x_{2t})=1$ and $P\left(2(\beta_{1}x_{1t}+\beta_{2}x_{2t})=\hbeta_{1t}x_{1t}+\hbeta_{2t}x_{2t}\right)\neq1$,
then we do not have $P\left[E(\Delta L_{t+1}\mid\mathcal{F}_{t})=0\right]=1$.
\end{prop}
Proposition \ref{prop: non-nested model} holds regardless of the
sample size, $m$, used to estimate the parameters $\hbeta_{it}$,
$i=1,2$. This again means that the MDS condition fails when $m$
is fixed and $n$ tends to infinity or when both $m$ and $n$ tend
to infinity.\footnote{Part 2 of Proposition \ref{prop: non-nested model} allows for any
estimator that uses information up to time $t$ and so is not limited
to the OLS estimator.}

\section{Practical Implications}

Proposition \ref{prop: master prop} has some interesting practical
implications. To see this, define the forecast error from a simple
equal-weighted average of forecasts:
\begin{equation}
\xi_{t+1}=y_{t+1}-(f_{1,t}+f_{2,t})/2.\label{eq:MDS variable}
\end{equation}
From Proposition \ref{prop: master prop}, the MDS null hypothesis
for the loss differential can equivalently be stated as $E(\xi_{t+1}\mid\mathcal{F}_{t})=0$,
motivating a test that is different from the GW test that is based
on the loss differential. 

A priori, it may not be obvious whether it is better, in a given sample,
to test $E(\Delta L_{t+1}\mid\mathcal{F}_{t})=0$ or $E(\xi_{t+1}\mid\mathcal{F}_{t})=0$.
Although such tests are equivalent, their finite sample power could
well be very different and practitioners will want to maximize power
in testing the MDS null. In some situations, an alternative test based
on $\xi_{t+1}$ might have better power properties than one based
on $\Delta L_{t+1}$. 

We next illustrate this point through an analytical example and an
empirical application.

\subsection{Power for AR(1) Process}

Following \citet{mccracken2020tests}, consider the following first-order
autoregressive DGP:
\begin{equation}
y_{t+1}=\rho y_{t}+\varepsilon_{t+1},
\end{equation}
where $\varepsilon_{t+1}$ is i.i.d with $E\varepsilon_{t+1}=0$ and
$E\varepsilon_{t+1}^{2}=\sigma^{2}$. Assume that $E|\varepsilon_{t+1}|^{3}$
is bounded. Let $f_{1,t}=y_{t}$ and $f_{2,t}=0$ so that $E(\Delta L_{t+1}\mid\mathcal{F}_{t})=0$
when $\rho=1/2$. To examine local power, let $\rho=1/2-T^{-1/2}c$,
where $c\in\mathbb{R}$ is a constant and $T$ is the sample sized
used for the test. To simplify notations, let $h_{t}=(f_{1,t}+f_{2,t})/2$
and consider the following test statistics: 
\begin{equation}
GW_{T}=\frac{T^{-1/2}\sum_{t=1}^{T}\Delta L_{t+1}h_{t}}{\sqrt{T^{-1}\sum_{t=1}^{T}\left(\Delta L_{t+1}h_{t}\right)^{2}}},
\end{equation}
and 
\begin{equation}
J_{T}=\frac{T^{-1/2}\sum_{t=1}^{T}\xi_{t+1}h_{t}}{\sqrt{T^{-1}\sum_{t=1}^{T}\left(\xi_{t+1}h_{t}\right)^{2}}}.
\end{equation}
Both test statistics use critical values from a standard Gaussian
distribution, $\Phi$. For a test of nominal size $\alpha$, we therefore
reject the null if $|GW_{T}|>cv_{\alpha}\equiv\Phi^{-1}(1-\alpha/2)$
and $|J_{T}|>cv_{\alpha}$ under the GW and $J$ tests, respectively.
The following result can be used to compute the power of the tests.
\begin{prop}
\label{prop: power comparison}Suppose $y_{t+1}$ is generated from
a first-order autoregressive process $y_{t+1}=\rho y_{t}+\varepsilon_{t+1}$.
Then
\[
\lim_{T\rightarrow\infty}P\left(|GW_{T}|>cv_{\alpha}\right)=P\left(\left|\frac{cEY^{3}}{\sqrt{EY^{4}}\sigma}+Z\right|>cv_{\alpha}\right),
\]
and 
\[
\lim_{T\rightarrow\infty}P\left(|J_{T}|>cv_{\alpha}\right)=P\left(\left|\frac{c\sqrt{EY^{2}}}{\sigma}+Z\right|>cv_{\alpha}\right),
\]
where $Z\sim N(0,1)$ and $Y$ follows the stationary distribution
of $y_{t}$. 
\end{prop}
This result implies that the local power of $J_{T}$ always exceeds
that of the $GW_{T}$ test for this example. To see this, notice that
by Holder's inequality, 
\[
|EY^{3}|\leq E|Y|^{3}\leq\sqrt{EY^{4}}\times\sqrt{EY^{2}},
\]
which implies that 
\[
\left|\frac{cEY^{3}}{\sqrt{EY^{4}}\sigma}\right|\leq\left|\frac{c\sqrt{EY^{2}}}{\sigma}\right|.
\]

\begin{figure}[H]
\caption{\label{fig: power}Power of GW and J tests as a function of $c$ when
$\varepsilon_{t}$ follows a mean-zero normal distribution.}

\centering{}\includegraphics[scale=0.5]{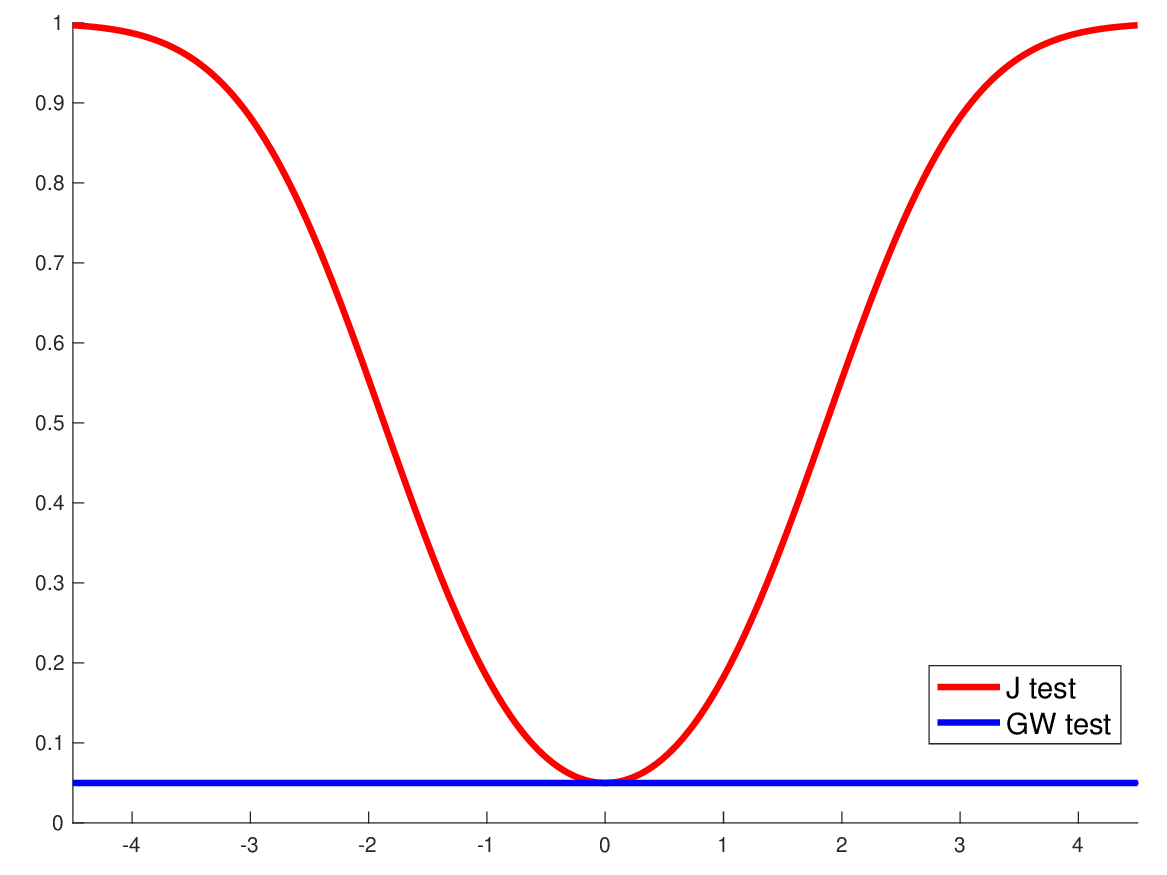}
\end{figure}

If $\varepsilon_{t}$ has a symmetric distribution, i.e., $\varepsilon_{t}$
and $-\varepsilon_{t}$ have the same distribution, then the test
based on $GW_{T}$ has no power at all because $Y$ has a symmetric
distribution and thus $EY^{3}=0$. For example, if $\varepsilon_{t}\sim N(0,\sigma^{2})$,
then $Y\sim N(0,\sigma^{2}/(1-\rho^{2}))$, which implies that $EY^{3}=0$
and $\lim_{T\rightarrow\infty}P\left(|GW_{T}|>cv_{\alpha}\right)=\alpha$
for any $c$. We illustrate this result in Figure \ref{fig: power}.\textcolor{red}{{} }

For this example, $\sqrt{EY^{2}}/\sigma=(1-\rho^{2})^{-1}$, which
means that the power of the test based on $J_{T}$ increases in $|\rho|$;
in other words, the power of this test is higher for more persistent
DGP's. 

It is worth emphasizing that the result in Proposition \ref{prop: power comparison}
hinges on using the equal-weighted forecast as our instrument, $h_{t}$.
The key point is that although $J_{T}$ and $GW_{T}$ are testing
the same MDS null for loss differences, they can have very different
power properties for a given choice of instrument.

\subsection{Empirical Application}

We next illustrate how the insights from our analysis can be used
to compare the accuracy of forecasts of two important economic variables,
namely US monthly inflation and GDP growth. Specifically, we obtain
data from the St. Louis Federal Reserve on the seasonally adjusted
consumer price index for all urban consumers (CPIAUCSL) from 1947:01
to 2022:07. From this we compute the monthly inflation rate. Next,
we consider the seasonally adjusted Gross Domestic Product (GDP),
again obtained from the St. Louis Federal Reserve Fred data base.
This is a quarterly series and runs from 1947Q1 through 2022Q2. Again
we compute the quarter-over-quarter growth rate and use this as our
dependent variable.

For both variables our benchmark models the conditional mean of the
dependent variable as a constant while the alternative model uses
an AR(4) process. Model parameters are estimated recursively using
a 10-year rolling window that gets updated as new data arrive.

As our first instrument, we use the lagged value of the equal-weighted
forecast error, i.e., $h_{t}=\xi_{t}=y_{t}-(f_{1,t-1}+f_{2,t-1})/2$.
For this case, we obtain test statistics $GW_{T}=1.45$ and $J_{T}=4.56$
for the inflation rate data and $GW_{T}=0.24$ and $J_{T}=2.27$ for
the GDP growth data. Hence, in both cases the $J$ test strongly rejects
the null while the GW test fails to reject.

The empirical results will of course depend on the chosen instrument.
To examine the robustness of our finding, we also consider using instead
the difference between the two forecasts as our instrument, setting
$h_{t}=f_{1,t-1}-f_{2,t-1}$. Using this instrument, we obtain test
statistics $GW_{T}=-1.09$ and $J_{T}=6.18$ for the inflation data
and $GW_{T}=-1.30$ and $J_{T}=1.22$ for the GDP growth data. In
this case, the $J$ test strongly rejects the null of equal conditional
predictive accuracy for the inflation data while the $GW$ test fails
to reject. Neither test rejects the null for the GDP growth rate data.

\section{Testing Equal Unconditional Expected Predictive Accuracy}

So far we have demonstrated that the null that the loss differential
follows an MDS generally, though not always, fails to hold conditional
on the information set used to generate the forecasts. One might wonder
what the GW test is actually testing when the MDS null fails. The
obvious candidate is the corresponding unconditional null:
\begin{equation}
H_{0}:\ E[\Delta L_{t+1}]=0.\label{eq:H0 unconditional}
\end{equation}

To examine whether we can use the GW approach to test the null in
(\ref{eq:H0 unconditional}), we separately consider cases with a
rolling and an expanding estimation window. 

\subsection{\label{subsec: Rolling-window GW}Rolling estimation window}

To see what happens with a rolling estimation window, consider the
following simple example: 
\begin{example}
\label{exa:2}Suppose that $y_{t+1}=c+\varepsilon_{t+1},$where $\varepsilon_{t}$
is iid with $E\varepsilon_{t}=0$, $E\varepsilon_{t}^{2}=1,$$E\varepsilon_{t}^{3}=\kappa_{1}$
and $E\varepsilon_{t}^{4}=\kappa_{2}$. We use $f_{1,t}=m^{-1}\sum_{s=t-m+1}^{t}y_{t}$
and $f_{2,t}=0$. Assuming squared error loss, we can simply choose
$c=m^{-1/2}$ to make the unconditional null hold (i.e., $E(\Delta L_{t+1})=0$). 
\end{example}
For the rolling window case, $m$ is fixed and $n$ tends to infinity.
Because $m$ is fixed, $\Delta L_{t+1}$ is weakly dependent and stationary.
However, the asymptotic distribution of the GW test statistic under
the unconditional null hypothesis is not $N(0,1)$. To obtain higher
power, GW prefer to use an estimator of the variance of the loss differences
that exploits the MDS property of loss differences under the null.
However, they also note in Comment 5 that one can use a HAC estimator
in situations with, e.g., positive autocorrelation in loss differences.
In practice, tests of the unconditional null in (\ref{eq:H0 unconditional})
are therefore often conducted using an approach similar to that adopted
by \citet{Diebold1995} which uses a consistent estimator of the long-run
variance of the loss differences and so accounts for any temporal
dependencies that may exist.

Using Example \ref{exa:2}, we next show that $J_{T}$ converges in
distribution to a normal distribution but with variance different
from one:
\begin{prop}
\label{prop: GW size distortion}In Example \ref{exa:2}, $GW_{T}$
converges in distribution to $N(0,V_{m})$, where 
\[
V_{m}=\frac{4m^{2}-4m^{3/2}\kappa_{1}+m(\kappa_{2}+3)}{8m^{2}+\kappa_{2}-1}.
\]
\end{prop}
From Proposition \ref{prop: GW size distortion}, the GW test introduces
size distortions asymptotically if $V_{m}>1$. This can easily be
the case for skewed distributions. For example, let $\log\xi_{t}\sim N(0,\sigma^{2})$
and $\varepsilon_{t}=-(\xi_{t}-E(\xi_{t}))/\sqrt{Var(\xi_{t})}$.
Then one can use simulations to see that $V_{m}>3$ for $\sigma=1.5$.
Moreover, the asymptotic variance can be arbitrarily close to 1/2
as the rolling window size $m$ increases, leading to an undersized
test. 
\begin{rem}
We can fix this issue by replacing the denominator in $GW_{T}$ with
the \citet{Newey1987a} or another heteroskedasticity and autocorrelation
consistent (HAC) estimator. This is in fact the procedure recommended
for the classical test proposed by \citet{Diebold1995}. The drawback
is that even if $\varepsilon_{t}$ is i.i.d, $\Delta L_{t+1}$ has
non-zero autocorrelation for at least $m$ lags. The performance of
Newey-West or other HAC estimates might not be satisfactory when the
serial dependence in the loss differentials does not decay fast enough.
\end{rem}

\begin{rem}
Another possibility is to use subsample t-tests similar to those proposed
by \citet{ibragimov2010t,ibragimov2016inference}. Thus, suppose we
divide $\{\Delta L_{t+1}\}_{t=m+1}^{m+n}$ into $K$ blocks and let
$\Delta\bar{L}^{(k)}$ denote the sample mean of $\Delta L_{t+1}$
in the $k$-th block, $k=1,...,K$. Consider the test statistic
\begin{equation}
S_{K}=\frac{K^{1/2}\overline{\Delta L}}{\sqrt{(K-1)^{-1}\sum_{k=1}^{K}(\Delta\bar{L}^{(k)}-\overline{\Delta L})^{2}}},\label{eq:Sk}
\end{equation}
where $\overline{\Delta L}=K^{-1}\sum_{k=1}^{K}\Delta\bar{L}^{(k)}$.
The limiting distribution of $S_{K}$ is the student $t$-distribution
with $K-1$ degrees of freedom. 
\end{rem}

\begin{rem}
It is well known that the outcome of tests of equal unconditional
predictive accuracy can be sensitive to the choice of the bandwidth
used to estimate the long-run variance of the loss differential, see
\citet{mccracken2020tests} and \citet{coroneo2020comparing}. Choosing
the right bandwidth is often challenging, and \citet{mccracken2020tests}
finds that very large bandwidths can be required for some data generating
processes. An advantage of the above subsample $t$-test is that it
does not require a well-defined long-run variance. For example, the
variance can change in a non-stationary manner, e.g. with structural
breaks, as long as it is bounded. 
\end{rem}
\begin{prop}
\label{prop: sub-t}Suppose that $\{\Delta L_{t+1}\}_{t=m+1}^{m+n}$
is stationary and $E\left[\Delta L_{t+1}\right]=0$. Assume that $E|\Delta L_{t+1}|^{r}$
is bounded for some $r>2$ and $\Delta L_{t+1}$ is strong mixing
of size $-r/(r-2)$. Then $S_{K}$ converges in distribution to the
student $t$-distribution with $K-1$ degrees of freedom. 
\end{prop}
The self-normalizing feature used to construct the test statistic
in (\ref{eq:Sk}) means that we do not need to explicitly compute
a HAC estimate, although we still need to choose $K$.\footnote{HAC estimates require us to choose the number of lags to include which,
in practice, can be quite complicated.} 

The key assumption needed for Proposition \ref{prop: sub-t} is stationarity
of the loss difference $\Delta L_{t+1}$. Conversely, the proof does
not require us to specify the functional form of the loss (MSE or
other loss) and allows for nonlinear models with general estimators
computed using a rolling window. 

\subsection{Monte Carlo simulation results}

Consider the setting in Section \ref{subsec: Rolling-window GW}.
We set $\log\xi_{t}\sim N(0,\sigma^{2})$ and $\varepsilon_{t}=-(\xi_{t}-E(\xi_{t}))/\sqrt{Var(\xi_{t})}$.
We report the size of three tests: the original GW test (GW), the
Diebold-Mariano test using Newey-West standard errors (DM)\footnote{The number of lags follows the ``textbook NW'' \citep{lazarus2018har}
choice and is set to $0.75T^{1/3}$. } and a subsample t-test $S_{K}$ with $K=2$ (Sub). All Monte Carlo
experiments are based on 10,000 random samples. To study the asymptotic
distribution of the tests, we set the sample size to be a large number
($n=20000$), but we also consider finite-sample performance in samples
with $n=100,200$ or 1,000 observations. 

Table \ref{tab:1} reports the results. First, consider the performance
of the tests in the very large sample ($n=20,000$). The last three
columns show that the original GW test tends to have an incorrect
size. For $\sigma=0.5$, the original GW test is oversized for small
$m$ ($m=3$) and undersized for large $m$ ($m=30$), while both
the DM and sub-sampling tests have approximately the right size. For
$\sigma=1.5$, we observe serious size distortions for the original
GW test which strongly over-rejects. Using Newey-West standard errors
improves the accuracy of the GW test but clearly fails to effectively
address the issue and this test remains heavily oversized. By far
the most accurate test is the subsample t-test of \citet{ibragimov2010t,ibragimov2016inference}
for which we only see a very small tendency to over-reject (e.g.,
6\% for a nominal size of 5\%). Similar results are seen in the finite
samples (columns 1-9) with the original GW and DM Newey-West test
statistics tending to over-reject, while the sub-sampling approach
is only modestly oversized. 

\begin{table}[H]
\caption{\label{tab:1}Rejection probability of a 5\% test under the null hypothesis}

\centering{}%
\begin{tabular}{ccccccccccccc}
 &  &  &  &  &  &  &  &  &  &  &  & \tabularnewline
 & \multicolumn{3}{c}{{\footnotesize{}$n=100$}} & \multicolumn{3}{c}{{\footnotesize{}$n=200$}} & \multicolumn{3}{c}{{\footnotesize{}$n=1000$}} & \multicolumn{3}{c}{{\footnotesize{}$n=20000$}}\tabularnewline
\hline 
 & {\footnotesize{}GW} & {\footnotesize{}DM} & {\footnotesize{}Sub} & {\footnotesize{}GW} & {\footnotesize{}DM} & {\footnotesize{}Sub} & {\footnotesize{}GW} & {\footnotesize{}DM} & {\footnotesize{}Sub} & {\footnotesize{}GW} & {\footnotesize{}DM} & {\footnotesize{}Sub}\tabularnewline
 &  &  &  &  &  &  &  &  &  &  &  & \tabularnewline
\hline 
{\footnotesize{}$m$} & \multicolumn{3}{c}{{\footnotesize{}$\sigma=0.5$}} & \multicolumn{3}{c}{{\footnotesize{}$\sigma=0.5$}} & \multicolumn{3}{c}{{\footnotesize{}$\sigma=0.5$}} & \multicolumn{3}{c}{{\footnotesize{}$\sigma=0.5$}}\tabularnewline
\hline 
{\footnotesize{}3} & {\footnotesize{}0.0915} & {\footnotesize{}0.0860} & {\footnotesize{}0.0465} & {\footnotesize{}0.0952} & {\footnotesize{}0.0792} & {\footnotesize{}0.0495} & {\footnotesize{}0.0895} & {\footnotesize{}0.0580} & {\footnotesize{}0.0468} & {\footnotesize{}0.0955} & {\footnotesize{}0.0530} & {\footnotesize{}0.0537}\tabularnewline
{\footnotesize{}5} & {\footnotesize{}0.0742} & {\footnotesize{}0.0799} & {\footnotesize{}0.0487} & {\footnotesize{}0.0737} & {\footnotesize{}0.0723} & {\footnotesize{}0.0498} & {\footnotesize{}0.0725} & {\footnotesize{}0.0598} & {\footnotesize{}0.0483} & {\footnotesize{}0.0755} & {\footnotesize{}0.0537} & {\footnotesize{}0.0521}\tabularnewline
{\footnotesize{}10} & {\footnotesize{}0.0545} & {\footnotesize{}0.0732} & {\footnotesize{}0.0524} & {\footnotesize{}0.0527} & {\footnotesize{}0.0616} & {\footnotesize{}0.0505} & {\footnotesize{}0.0543} & {\footnotesize{}0.0595} & {\footnotesize{}0.0517} & {\footnotesize{}0.0554} & {\footnotesize{}0.0541} & {\footnotesize{}0.0496}\tabularnewline
{\footnotesize{}30} & {\footnotesize{}0.0430} & {\footnotesize{}0.0604} & {\footnotesize{}0.0543} & {\footnotesize{}0.0378} & {\footnotesize{}0.0500} & {\footnotesize{}0.0500} & {\footnotesize{}0.0381} & {\footnotesize{}0.0452} & {\footnotesize{}0.0502} & {\footnotesize{}0.0349} & {\footnotesize{}0.0435} & {\footnotesize{}0.0484}\tabularnewline
 &  &  &  &  &  &  &  &  &  &  &  & \tabularnewline
\hline 
{\footnotesize{}$m$} & \multicolumn{3}{c}{{\footnotesize{}$\sigma=1$}} & \multicolumn{3}{c}{{\footnotesize{}$\sigma=1$}} & \multicolumn{3}{c}{{\footnotesize{}$\sigma=1$}} & \multicolumn{3}{c}{{\footnotesize{}$\sigma=1$}}\tabularnewline
\hline 
{\footnotesize{}3} & {\footnotesize{}0.2593} & {\footnotesize{}0.2022} & {\footnotesize{}0.0585} & {\footnotesize{}0.2568} & {\footnotesize{}0.1748} & {\footnotesize{}0.0554} & {\footnotesize{}0.2489} & {\footnotesize{}0.1217} & {\footnotesize{}0.0543} & {\footnotesize{}0.2383} & {\footnotesize{}0.0725} & {\footnotesize{}0.0517}\tabularnewline
{\footnotesize{}5} & {\footnotesize{}0.2282} & {\footnotesize{}0.1883} & {\footnotesize{}0.0564} & {\footnotesize{}0.2364} & {\footnotesize{}0.1772} & {\footnotesize{}0.0590} & {\footnotesize{}0.2368} & {\footnotesize{}0.1179} & {\footnotesize{}0.0461} & {\footnotesize{}0.2589} & {\footnotesize{}0.0734} & {\footnotesize{}0.0489}\tabularnewline
{\footnotesize{}10} & {\footnotesize{}0.1680} & {\footnotesize{}0.1573} & {\footnotesize{}0.0510} & {\footnotesize{}0.1708} & {\footnotesize{}0.1451} & {\footnotesize{}0.0506} & {\footnotesize{}0.1928} & {\footnotesize{}0.1255} & {\footnotesize{}0.0508} & {\footnotesize{}0.2053} & {\footnotesize{}0.0762} & {\footnotesize{}0.0508}\tabularnewline
{\footnotesize{}30} & {\footnotesize{}0.1029} & {\footnotesize{}0.1182} & {\footnotesize{}0.0505} & {\footnotesize{}0.1030} & {\footnotesize{}0.1059} & {\footnotesize{}0.0458} & {\footnotesize{}0.1037} & {\footnotesize{}0.0946} & {\footnotesize{}0.0482} & {\footnotesize{}0.1213} & {\footnotesize{}0.0854} & {\footnotesize{}0.0488}\tabularnewline
 &  &  &  &  &  &  &  &  &  &  &  & \tabularnewline
\hline 
{\footnotesize{}$m$} & \multicolumn{3}{c}{{\footnotesize{}$\sigma=1.5$}} & \multicolumn{3}{c}{{\footnotesize{}$\sigma=1.5$}} & \multicolumn{3}{c}{{\footnotesize{}$\sigma=1.5$}} & \multicolumn{3}{c}{{\footnotesize{}$\sigma=1.5$}}\tabularnewline
\hline 
{\footnotesize{}3} & {\footnotesize{}0.5324} & {\footnotesize{}0.4635} & {\footnotesize{}0.1246} & {\footnotesize{}0.5196} & {\footnotesize{}0.4268} & {\footnotesize{}0.1084} & {\footnotesize{}0.4942} & {\footnotesize{}0.3481} & {\footnotesize{}0.0966} & {\footnotesize{}0.4182} & {\footnotesize{}0.2293} & {\footnotesize{}0.0667}\tabularnewline
{\footnotesize{}5} & {\footnotesize{}0.5028} & {\footnotesize{}0.4274} & {\footnotesize{}0.1091} & {\footnotesize{}0.5166} & {\footnotesize{}0.4118} & {\footnotesize{}0.0969} & {\footnotesize{}0.5301} & {\footnotesize{}0.3362} & {\footnotesize{}0.0896} & {\footnotesize{}0.5048} & {\footnotesize{}0.2195} & {\footnotesize{}0.0640}\tabularnewline
{\footnotesize{}10} & {\footnotesize{}0.4241} & {\footnotesize{}0.3819} & {\footnotesize{}0.0875} & {\footnotesize{}0.4497} & {\footnotesize{}0.3754} & {\footnotesize{}0.0867} & {\footnotesize{}0.5052} & {\footnotesize{}0.3362} & {\footnotesize{}0.0788} & {\footnotesize{}0.5620} & {\footnotesize{}0.2218} & {\footnotesize{}0.0605}\tabularnewline
{\footnotesize{}30} & {\footnotesize{}0.2698} & {\footnotesize{}0.2718} & {\footnotesize{}0.0673} & {\footnotesize{}0.2979} & {\footnotesize{}0.2828} & {\footnotesize{}0.0721} & {\footnotesize{}0.3523} & {\footnotesize{}0.2894} & {\footnotesize{}0.0656} & {\footnotesize{}0.4707} & {\footnotesize{}0.2357} & {\footnotesize{}0.0639}\tabularnewline
 &  &  &  &  &  &  &  &  &  &  &  & \tabularnewline
\hline 
\end{tabular}
\end{table}

\subsection{Expanding estimation window}

Proposition \ref{prop: impossibility rolling window} demonstrates
the implausibility of the MDS condition with a rolling window estimation
scheme. However, one might wonder whether testing equal predictive
accuracy would be easier if one adopts an expanding estimation window.
We next examine this point through a simple example: 
\begin{example}
\label{exa:1}Consider the following DGP:
\end{example}
\[
y_{t}=\varepsilon_{t},
\]
where $\{\varepsilon_{t}\}_{t=1}^{T}$ is i.i.d with $E\varepsilon_{t}=0$
and $E\varepsilon_{t}^{2}=1$. Let $\mathcal{F}_{t}$ denote the $\sigma$-algebra
generated by $\{\varepsilon_{1},...,\varepsilon_{t}\}$.

Consider an expanding window estimation scheme under which the forecast
for $y_{t+1}$ at time $t$ is $f_{1,t}=t^{-1}\sum_{s=1}^{t}y_{s}$,
where $t\geq m$ with $m\rightarrow\infty$ and $m/T\rightarrow\lambda$
for $\lambda\in(0,1)$. We compare this forecast with the simple prediction
$f_{2,t}=t^{-1/2}$. Clearly, both $f_{1,t}$ and $f_{2,t}$ have
a vanishing bias for $E(y_{t+1})=0$. Under squared error loss, one
can easily verify that $E(\Delta L_{t+1})=0$ and $E(\Delta L_{t+1}\mid\mathcal{F}_{t})=f_{1,t}^{2}-f_{1,t}^{2}$.
Since the bias vanishes with the sample size as the estimation window
expands, this example is similar to the local-to-zero setting considered
by \citet{clark2015nested}.
\begin{prop}
\label{prop: example 1}In Example \ref{exa:1}, the limiting distribution
of the GW statistic $GW_{T}$ in (\ref{eq: GW stat}) is 
\end{prop}
\begin{equation}
\frac{\int_{\lambda}^{1}(u^{-2}B^{2}(u)-u^{-1})du}{2\sqrt{\int_{\lambda}^{1}\left[u^{-1/2}-u^{-1}B(u)\right]^{2}du}}-\frac{\int_{\lambda}^{1}\left[u^{-1/2}-u^{-1}B(u)\right]dB(u)}{\sqrt{\int_{\lambda}^{1}\left[u^{-1/2}-u^{-1}B(u)\right]^{2}du}},\label{eq: limiting dist expanding}
\end{equation}
where $B(\cdot)$ is a standard Brownian motion. 

The second term in (\ref{eq: limiting dist expanding}) has a $N(0,1)$
distribution: 
\[
\frac{\int_{\lambda}^{1}\left[u^{-1/2}-u^{-1}B(u)\right]dB(u)}{\sqrt{\int_{\lambda}^{1}\left[u^{-1/2}-u^{-1}B(u)\right]^{2}du}}\sim N(0,1)\qquad\forall\lambda\in(0,1).
\]

Conversely, the first term in (\ref{eq: limiting dist expanding})
has a non-standard distribution and so the limiting distribution of
$GW_{T}$ is also non-standard.\footnote{This is similar to the result for the MSE-t test in Theorem 3.2 of
\citet{Clark2005a}.} 

In Table \ref{tab:0}, we simulate the limiting distribution in (\ref{eq: limiting dist expanding})
and tabulate the 95\% quantile of the absolute value limiting distribution
for various values of $\lambda$. We also record the asymptotic null
rejection probability if we simply choose 1.96 as the critical value
(the standard normal limiting distribution stated in GW). We observe
substantial size distortions if $\lambda$ is small, even in the limit.
Hence, in practice, if the expanding window starts early in the sample,
we would falsely reject the null hypothesis too often. 

\begin{table}[H]
\caption{\label{tab:0}Quantiles of the limiting distribution of $|GW_{T}|$
in (\ref{eq: limiting dist expanding})}

\centering{}%
\begin{tabular}{ccc}
$\lambda$ & 95\% quantile & Size if use 1.96\tabularnewline
0.05 & 3.993 & 0.247\tabularnewline
0.10 & 3.769 & 0.226\tabularnewline
0.15 & 3.573 & 0.215\tabularnewline
0.20 & 3.389 & 0.203\tabularnewline
0.25 & 3.250 & 0.196\tabularnewline
0.30 & 3.103 & 0.188\tabularnewline
0.35 & 2.981 & 0.181\tabularnewline
0.40 & 2.880 & 0.175\tabularnewline
0.45 & 2.781 & 0.166\tabularnewline
0.50 & 2.697 & 0.158\tabularnewline
0.55 & 2.598 & 0.147\tabularnewline
0.60 & 2.534 & 0.135\tabularnewline
0.65 & 2.445 & 0.122\tabularnewline
0.70 & 2.379 & 0.111\tabularnewline
0.75 & 2.307 & 0.099\tabularnewline
0.80 & 2.238 & 0.089\tabularnewline
0.85 & 2.190 & 0.081\tabularnewline
0.90 & 2.108 & 0.070\tabularnewline
0.95 & 2.057 & 0.062\tabularnewline
0.99 & 1.992 & 0.054\tabularnewline
 &  & \tabularnewline
\end{tabular}
\end{table}

\section{Conclusion}

Economic forecasts feature prominently in governments' decisions on
fiscal policy, central banks' monetary policy, households' consumption
and investment decisions and companies' hiring and capital expenditure
choices, so it is important to be able to tell if one forecast can
be expected to be more accurate than an alternative forecast. In an
influential and innovative paper, \citet{giacomini2006tests} develop
methods for testing the null hypothesis that two forecasts have identical
conditionally expected loss. Equivalently, their null is that the
loss differential follows a martingale difference sequence. They use
this null to construct a test statistic that does not require Newey-West
HAC type adjustments for serial correlation in loss differentials. 

The Giacomini-White approach has been used extensively in empirical
work as it provides a way to formally compare the accuracy of economic
forecasts in situations that are challenging for other tests such
as the case with nested models. It turns out that the null that the
loss differential is a martingale difference sequence is quite restrictive.
We establish that a necessary and sufficient condition for the conditionally
expected loss differential of two forecasts to follow an MDS process
is that the conditional expectation of the outcome is a simple average
of the forecasts. When the underlying forecasts contain parameter
estimation errors, this means that the conditional mean of the outcome
depends on past estimation errors. One can construct examples where
this condition is valid but in many settings of interest to economic
forecasters this condition seems hard to justify.

\appendix

\section{Proofs}
\begin{proof}[\textbf{Proof of Proposition \ref{prop: master prop}}]
Notice that 
\begin{align*}
\Delta L_{t+1} & =(y_{t+1}-f_{1,t})^{2}-(y_{t+1}-f_{2,t})^{2}\\
 & =f_{1,t}^{2}-f_{2,t}^{2}-2y_{t+1}(f_{1,t}-f_{2,t})\\
 & =(f_{1,t}+f_{2,t}-2y_{t+1})(f_{1,t}-f_{2,t}).
\end{align*}

Since $f_{1,t}$ and $f_{2,t}$ are both $\mathcal{F}_{t}$-measurable,
it follows that 
\[
E(\Delta L_{t+1}\mid\mathcal{F}_{t})=\left[f_{1,t}+f_{2,t}-2E(y_{t+1}\mid\mathcal{F}_{t})\right](f_{1,t}-f_{2,t}).
\]

Since $f_{1,t}-f_{2,t}\neq0$ almost surely, $E(\Delta L_{t+1}\mid\mathcal{F}_{t})=0$
if and only if $f_{1,t}+f_{2,t}-2E(y_{t+1}\mid\mathcal{F}_{t})=0$,
which implies $E(y_{t+1}\mid\mathcal{F}_{t})=(f_{1,t}+f_{2,t})/2$. 
\end{proof}
\begin{proof}[\textbf{Proof of Proposition \ref{prop: impossibility rolling window}}]
We proceed by contradiction. Suppose the MDS condition holds, i.e.,
$E(\Delta L_{t+1}\mid\mathcal{F}_{t})=0$ with probability one. Then
by Proposition \ref{prop: master prop}, $E(y_{t+1}\mid\mathcal{F}_{t})=(f_{1,t}+f_{2,t})/2$
almost surely. By the law of iterated expectation, we have 
\[
E(y_{t+1}\mid\mathcal{F}_{t-m:t})=E\left[E(y_{t+1}\mid\mathcal{F}_{t})\mid\mathcal{F}_{t-m:t}\right]=E\left[(f_{1,t}+f_{2,t})/2\mid\mathcal{F}_{t-m:t}\right]=(f_{1,t}+f_{2,t})/2.
\]

Therefore, $E(y_{t+1}\mid\mathcal{F}_{t-m:t})=(f_{1,t}+f_{2,t})/2=E(y_{t+1}\mid\mathcal{F}_{t})$
almost surely. This contradicts the assumption of $P(E(y_{t+1}\mid\mathcal{F}_{t})=E(y_{t+1}\mid\mathcal{F}_{t-m:t}))<1$.
The desired result follows. 
\end{proof}

\begin{proof}[\textbf{Proof of Lemma \ref{lem: AR example}}]
We proceed by contradiction. Suppose that $E(\Delta L_{t+1}\mid\mathcal{F}_{t})=0$.
Let $\hbeta_{t}=(\hbeta_{t,1}+\hbeta_{t,2})/2$ and $\rho=(\rho_{1},...,\rho_{q})'$.
Then by Proposition \ref{prop: master prop}, we have that $E(Y_{t+1}\mid\mathcal{F}_{t})=(f_{1,t}+f_{2,t})/2=Y_{t-q:t}'\hbeta_{t}$.
On the other hand, since $Y_{t+1}=\sum_{j=1}^{q}\rho_{j}Y_{t+1-j}+\varepsilon_{t+1}$
and $\varepsilon_{t+1}$ is i.i.d with $E(\varepsilon_{t+1})=0$,
we have that $E(Y_{t+1}\mid\mathcal{F}_{t})=Y_{t-q:t}'\rho$. Hence,
we have $Y_{t-q:t}'(\hbeta_{t}-\rho)=0$ almost surely. Therefore,
${\rm Var}\left(Y_{t-q:t}'(\hbeta_{t}-\rho)\mid Y_{t-q:t}\right)=0$
almost surely.

However, note that 
\[
{\rm Var}\left(Y_{t-q:t}'(\hbeta_{t}-\rho)\mid Y_{t-q:t}\right)=Y_{t-q:t}'\left[{\rm Var}(\hat{\beta}_{t,1}+\hbeta_{t,2}\mid Y_{t-q:t})\right]Y_{t-q:t}.
\]

Let $\mathcal{A}$ denote the event on which ${\rm Var}(\hat{\beta}_{t,1}+\hbeta_{t,2}\mid Y_{t-q:t})$
is positive definite. Since ${\rm Var}\left(Y_{t-q:t}'(\hbeta_{t}-\rho)\mid Y_{t-q:t}\right)=0$
almost surely, it follows that on the event $\mathcal{A}$, $Y_{t-q:t}=0$.
Thus, $P(Y_{t-q:t}=0)\geq P(\mathcal{A})$. By assumption, $P(\mathcal{A})>0$,
which means that $P(Y_{t-q:t}=0)>0$. However, this is impossible
because by assumption $Y_{t-q:t}$ has a density with respect to the
Lebesgue measure on $\mathbb{R}^{q}$ and thus $P(Y_{t-q:t}=0)=0$.
The proof is complete. 
\end{proof}
\begin{proof}[\textbf{Proof of Proposition \ref{prop: power comparison}}]
We prove the two claims in two steps.

\textbf{Step 1:} show the result for $GW_{T}$

By $y_{t+1}=(1/2-T^{-1/2}c)y_{t}+\varepsilon_{t+1}$, we have 
\begin{align*}
\Delta L_{t+1} & =(y_{t+1}-f_{1,t})^{2}-(y_{t+1}-f_{2,t})^{2}\\
 & =(y_{t+1}-y_{t})^{2}-(y_{t+1}-0)^{2}=2T^{-1/2}cy_{t}^{2}-2y_{t}\varepsilon_{t+1}.
\end{align*}

By $h_{t}=(f_{1,t}+f_{2,t})/2=y_{t}/2$, we have that $\Delta L_{t+1}h_{t}=2T^{-1/2}cy_{t}^{3}-2y_{t}^{2}\varepsilon_{t+1}$.
Let $Z_{t+1}=\Delta L_{t+1}h_{t}-E[\Delta L_{t+1}h_{t}\mid\mathcal{F}_{t}]$.
Clearly, $\{Z_{t},\mathcal{F}_{t}\}$ is an MDS and $E[\Delta L_{t+1}h_{t}\mid\mathcal{F}_{t}]=2n^{-1/2}cy_{t}^{3}$.
Hence, 
\[
Z_{t+1}=-2y_{t}\varepsilon_{t+1}
\]
 and 
\[
EZ_{t+1}^{2}=4\sigma^{2}EY_{t}^{2}.
\]

Since $Z_{t}$ is an MDS, we have 

\[
{\rm Var}\left(T^{-1/2}\sum_{t=1}^{n}Z_{t+1}\right)=T^{-1}\sum_{t=1}^{T}E(\Delta L_{t+1}h_{t})^{2}=4\sigma^{2}EY_{t}^{2}.
\]

By Theorem 3.35 of \citet{White2001}, $Z_{t+1}$ is stationary and
ergodic. By Theorem 3.34 therein, $T^{-1}\sum_{t=1}^{T}Z_{t+1}^{2}=4\sigma^{2}EY_{t}^{2}+o_{P}(1)$.
By Corollary 5.26 therein, we have 
\[
\frac{T^{-1/2}\sum_{t=1}^{T}Z_{t+1}}{2\sigma\sqrt{EY_{t}^{2}}}\rightarrow^{d}N(0,1).
\]

In other words, we have $T^{-1/2}\sum_{i=1}^{T}Z_{t+1}\rightarrow^{d}N(0,4\sigma^{2}EY_{t}^{2})$.
Again, by Theorems 3.34 and 3.35 of \citet{White2001}, we have 
\[
T^{-1/2}\sum_{t=1}^{T}E[\Delta L_{t+1}h_{t}\mid\mathcal{F}_{t}]=2cT^{-1}\sum_{i=1}^{T}y_{t}^{3}=2cEY_{t}^{3}+o_{P}(1).
\]

Therefore, 
\begin{equation}
T^{-1/2}\sum_{t=1}^{T}\Delta L_{t+1}h_{t}\rightarrow^{d}N(2cEY_{t}^{3},4\sigma^{2}EY_{t}^{2}).\label{eq: prop power eq 1}
\end{equation}

On the other hand, by $Z_{t+1}=\Delta L_{t+1}h_{t}-2n^{-1/2}cy_{t}^{3}$,
we have that 
\begin{align*}
\left|\sqrt{T^{-1}\sum_{t=1}^{T}Z_{t+1}^{2}}-\sqrt{T^{-1}\sum_{t=1}^{T}(\Delta L_{t+1}h_{t})^{2}}\right| & \leq\sqrt{T^{-1}\sum_{t=1}^{T}(Z_{t+1}-\Delta L_{t+1}h_{t})^{2}}\\
 & =\sqrt{T^{-1}\sum_{t=1}^{T}(2T^{-1/2}cy_{t}^{3})^{2}}=O_{P}(T^{-1})=o_{P}(1).
\end{align*}

Since $T^{-1}\sum_{t=1}^{T}Z_{t+1}^{2}=4\sigma^{2}EY_{t}^{2}+o_{P}(1)$,
we have 
\begin{equation}
T^{-1}\sum_{t=1}^{T}(\Delta L_{t+1}h_{t})^{2}=4\sigma^{2}EY_{t}^{2}+o_{P}(1).\label{eq: prop power eq 2}
\end{equation}

Combining (\ref{eq: prop power eq 1}) and (\ref{eq: prop power eq 2}),
we have 
\[
GW_{T}=\frac{T^{-1/2}\sum_{t=1}^{T}\Delta L_{t+1}h_{t}}{\sqrt{T^{-1}\sum_{t=1}^{T}(\Delta L_{t+1}h_{t})^{2}}}\rightarrow^{d}N\left(\frac{cEY_{t}^{3}}{\sigma\sqrt{EY_{t}^{2}}},1\right).
\]

\textbf{Step 2:} show the result for $J_{T}$. 

The result for $J_{T}$ follows by an analogous argument. We provide
a brief proof and point out the difference. Again, define $Z_{t+1}=\xi_{t+1}h_{t}-E[\xi_{t+1}h_{t}\mid\mathcal{F}_{t}]$,
where 
\[
\xi_{t+1}h_{t}=[y_{t+1}-y_{t}/2]\cdot y_{t}/2=\frac{1}{2}y_{t}\varepsilon_{t+1}+\frac{1}{2}n^{-1/2}cy_{t}^{2}
\]
and $E[\xi_{t+1}h_{t}\mid\mathcal{F}_{t}]=T^{-1/2}cy_{t}^{2}/2$.
Notice that besides a constant factor, the difference from Step 1
is that we now have $y_{t}^{2}$ rather than $y_{t}^{3}$. 

By essentially the same argument, we have 
\begin{equation}
T^{-1/2}\sum_{t=1}^{T}Z_{t+1}\rightarrow^{d}N(0,\sigma^{2}EY_{t}^{2}/4)\label{eq: prop power eq 3}
\end{equation}
and 
\begin{equation}
T^{-1/2}\sum_{t=1}^{T}E[\xi_{t+1}h_{t}\mid\mathcal{F}_{t}]=\frac{1}{2}cT^{-1}\sum_{i=1}^{T}y_{t}^{2}=\frac{1}{2}cEY_{t}^{2}+o_{P}(1).\label{eq: prop power eq 3.5}
\end{equation}
as well as
\begin{equation}
T^{-1}\sum_{t=1}^{T}(\xi_{t+1}h_{t})^{2}=\sigma^{2}EY_{t}^{2}/4+o_{P}(1).\label{eq: prop power eq 4}
\end{equation}

By (\ref{eq: prop power eq 3}) and (\ref{eq: prop power eq 3.5}),
we have that 
\[
T^{-1/2}\sum_{t=1}^{T}\xi_{t+1}h_{t}=T^{-1/2}\sum_{t=1}^{T}Z_{t+1}+T^{-1/2}\sum_{t=1}^{T}E[\xi_{t+1}h_{t}\mid\mathcal{F}_{t}]\rightarrow^{d}N(cEY_{t}^{2}/2,\sigma^{2}EY_{t}^{2}/4).
\]

Then by (\ref{eq: prop power eq 4}), we have that 
\[
J_{T}=\frac{T^{-1/2}\sum_{t=1}^{T}\xi_{t+1}h_{t}}{\sqrt{T^{-1}\sum_{t=1}^{T}(\xi_{t+1}h_{t})^{2}}}\rightarrow^{d}N\left(\frac{c\sqrt{EY_{t}^{2}}}{\sigma},1\right).
\]

The proof is complete. 
\end{proof}
\begin{proof}[\textbf{Proof of Proposition \ref{lem: size DGP}}]
To prove this result, we compute the mean squared error for the individual
models. To this end, define $y_{(t)}=(y_{2},\cdots,y_{t})'\in\mathbb{R}^{t}$
and $\varepsilon_{(t)}=(\varepsilon_{2},\cdots,\varepsilon_{t})'\in\mathbb{R}^{t}$.
For the small model, $y_{t+1}-f_{1,t}=c+(\beta-\hat{\theta}_{t})x_{t}+\varepsilon_{t+1}$.
Since $\hat{\theta}_{t}=x_{(t-1)}^{\prime}y_{(t)}/(x_{(t-1)}^{\prime}x_{(t-1)})$
and $y_{(t)}=c1_{t-1}+x_{(t-1)}\beta+\varepsilon_{(t)}$, we have
\[
y_{t+1}-f_{1,t}=c\left(1-\frac{x_{(t-1)}^{\prime}1_{t-1}}{x_{(t-1)}^{\prime}x_{(t-1)}}x_{t}\right)-\frac{x_{(t-1)}^{\prime}\varepsilon_{(t)}}{x_{(t-1)}^{\prime}x_{(t-1)}}x_{t}+\varepsilon_{t+1}.
\]
Simple computations yield 
\begin{equation}
E\left(y_{t+1}-f_{1,t}\right)^{2}=c^{2}\left(1-\frac{x_{(t-1)}^{\prime}1_{t-1}}{x_{(t-1)}^{\prime}x_{(t-1)}}x_{t}\right)^{2}+\sigma^{2}\left(1+\frac{x_{t}^{2}}{x_{(t-1)}^{\prime}x_{(t-1)}}\right).\label{eq: size DGP eq 1}
\end{equation}
For the big model, $y_{t+1}-f_{2,t}=(c-\hat{c}_{t})+(\beta-\hat{\beta}_{t})x_{t}+\varepsilon_{t+1}=\varepsilon_{t+1}-X_{t}^{\prime}\begin{pmatrix}\hat{c}_{t}-c\\
\hat{\beta}_{t}-\beta
\end{pmatrix}$. Since $\begin{pmatrix}\hat{c}_{t}-c\\
\hat{\beta}_{t}-\beta
\end{pmatrix}=\left(X_{(t-1)}^{\prime}X_{(t-1)}\right)^{-1}X_{(t-1)}^{\prime}\varepsilon_{(t)}$, we have $y_{t+1}-f_{2,t}=\varepsilon_{t+1}-X_{t-1}^{\prime}\left(X_{(t-1)}^{\prime}X_{(t-1)}\right)^{-1}X_{(t-1)}^{\prime}\varepsilon_{(t)}$.
By simple computations, we obtain 
\begin{equation}
E\left(y_{t+1}-f_{2,t}\right)^{2}=\left(1+X_{t-1}^{\prime}\left(X_{(t-1)}^{\prime}X_{(t-1)}\right)^{-1}X_{t-1}\right)\sigma^{2}.\label{eq: size DGP eq 2}
\end{equation}
The first result follows by setting $n^{-1}\sum_{t=m+1}^{m+n}E\left(y_{t+1}-f_{1,t}\right)^{2}=n^{-1}\sum_{t=m+1}^{m+n}E\left(y_{t+1}-f_{2,t}\right)^{2}$
and using (\ref{eq: size DGP eq 1}) and (\ref{eq: size DGP eq 2}).

We now show the second result. Let $\mathcal{F}_{t}$ be the $\sigma$-algebra
generated by $(\varepsilon_{1},...,\varepsilon_{t})$ and notice that
\[
E\left[\left(y_{t+1}-f_{2,t}\right)^{2}\mid\mathcal{F}_{t}\right]=X_{t-1}^{\prime}\left(X_{(t-1)}^{\prime}X_{(t-1)}\right)^{-1}X_{(t-1)}^{\prime}\varepsilon_{(t)}\varepsilon_{(t)}^{\prime}X_{(t-1)}\left(X_{(t-1)}^{\prime}X_{(t-1)}\right)^{-1}X_{t-1}+\sigma^{2},
\]
and 
\[
E\left[\left(y_{t+1}-f_{1,t}\right)^{2}\mid\mathcal{F}_{t}\right]=\left[c\left(1-\frac{x_{(t-1)}^{\prime}1_{t}}{x_{(t-1)}^{\prime}x_{(t-1)}}x_{t}\right)-\frac{x_{(t-1)}^{\prime}\varepsilon_{(t)}}{x_{(t-1)}^{\prime}x_{(t-1)}}x_{t}\right]^{2}+\sigma^{2}.
\]

Therefore, 
\begin{multline*}
E\left[\Delta L_{t+1}\mid\mathcal{F}_{t}\right]=X_{t-1}^{\prime}\left(X_{(t-1)}^{\prime}X_{(t-1)}\right)^{-1}X_{(t-1)}^{\prime}\varepsilon_{(t)}\varepsilon_{(t)}^{\prime}X_{(t-1)}\left(X_{(t-1)}^{\prime}X_{(t-1)}\right)^{-1}X_{t-1}\\
-\left[c\left(1-\frac{x_{(t-1)}^{\prime}1_{t-1}}{x_{(t-1)}^{\prime}x_{(t-1)}}x_{t}\right)-\frac{x_{(t-1)}^{\prime}\varepsilon_{(t)}}{x_{(t-1)}^{\prime}x_{(t-1)}}x_{t}\right]^{2}.
\end{multline*}

Since $X_{(t-1)}$ contains a column of $1_{t-1}$, there is always
a term containing $1_{t-1}^{\prime}\varepsilon_{(t)}$ that cannot
be canceled in the above equation. Therefore, it is not possible that
with probability one, $E[\Delta L_{t+1}\mid\mathcal{F}_{t}]=0$. Hence,
$\Delta L_{t+1}$ is not an MDS. 
\end{proof}

\begin{proof}[\textbf{Proof of Proposition \ref{prop: non-nested model}}]
We proceed in two steps in which we verify the result in the absence
and presence of estimation errors.

\textbf{Step 1:} First, ignore estimation errors. 

We proceed by contradiction. Suppose that $E(\Delta L_{t+1}\mid\mathcal{F}_{t})=0$
almost surely. In this case, $f_{1,t}=\beta_{1}x_{1t}$ and $f_{2,t}=\beta_{2}x_{2t}$.
By Proposition \ref{prop: master prop}, we have 
\[
E(y_{t+1}\mid\mathcal{F}_{t})=(f_{1,t}+f_{2,t})/2=(\beta_{1}x_{1t}+\beta_{2}x_{2t})/2.
\]

However, from the DGP $y_{t+1}=\beta_{1}x_{1t}+\beta_{2}x_{2t}+\varepsilon_{t+1}$
with $E(\varepsilon_{t+1}\mid\mathcal{F}_{t})=0$, we have 
\[
E(y_{t+1}\mid\mathcal{F}_{t})=\beta_{1}x_{1t}+\beta_{2}x_{2t}.
\]

It follows that $\beta_{1}x_{1t}+\beta_{2}x_{2t}=0$ almost surely.
This contradicts the assumption, from which the result follows.

\textbf{Step 2:} Next, consider parameter estimation errors. 

We proceed by contradiction. Suppose that $E(\Delta L_{t+1}\mid\mathcal{F}_{t})=0$
almost surely. In this case, $f_{1,t}=\hbeta_{1t}x_{1t}$ and $f_{2,t}=\hbeta_{2t}x_{2t}$,
where $\hbeta_{1t}$ and $\hbeta_{2t}$ are OLS estimates using information
in $\mathcal{F}_{t}$. By Proposition \ref{prop: master prop}, we
have 
\[
E(y_{t+1}\mid\mathcal{F}_{t})=(f_{1,t}+f_{2,t})/2=(\hbeta_{1t}x_{1t}+\hbeta_{2t}x_{2t})/2.
\]

On the other hand, we have $E(y_{t+1}\mid\mathcal{F}_{t})=\beta_{1}x_{1t}+\beta_{2}x_{2t}$.
This means that $2(\beta_{1}x_{1t}+\beta_{2}x_{2t})=\hbeta_{1t}x_{1t}+\hbeta_{2t}x_{2t}$
almost surely. This contradicts the assumption, from which the result
follows.
\end{proof}

\begin{proof}[\textbf{Proof of Proposition \ref{prop: GW size distortion}}]
Let $\beps_{t}=m^{-1}\sum_{s=t-m+1}^{t}\varepsilon_{s}$. Then $y_{t+1}-f_{1,t}=\varepsilon_{t+1}-\beps_{t}$
and $y_{t+1}-f_{2,t}=c+\varepsilon_{t+1}$. Hence, 
\[
\Delta L_{t+1}=(y_{t+1}-f_{1,t})^{2}-(y_{t+1}-f_{2,t})^{2}=\beps_{t}^{2}-c^{2}-2\varepsilon_{t+1}(c+\beps_{t}).
\]

Since $m$ is fixed, $\Delta L_{t+1}$ is stationary and weakly dependent;
in fact, $\Delta L_{t+1+d}$ and $\Delta L_{t+1}$ are independent
for $d\geq m+1$. We next compute the autocovariances, $\gamma_{d}=E\Delta L_{t+1+d}\Delta L_{t+1}$.
Clearly, $\gamma_{d}=0$ for $d>m$, so we can focus on $d\leq m$.
Since $\varepsilon_{t}$ is iid with mean zero and $c^{2}=m^{-1}$,
we observe that 
\begin{align}
\gamma_{d} & =E\Delta L_{t+1+d}\Delta L_{t+1}\nonumber \\
 & =E\left(\beps_{t+d}^{2}-c^{2}-2\varepsilon_{t+d+1}(c+\beps_{t+d})\right)\left(\beps_{t}^{2}-c^{2}-2\varepsilon_{t+1}(c+\beps_{t})\right)\nonumber \\
 & =E(\beps_{t+d}^{2}-c^{2})(\beps_{t}^{2}-c^{2})-2E(\beps_{t+d}^{2}-c^{2})\varepsilon_{t+1}(c+\beps_{t})\nonumber \\
 & =E(\beps_{t+d}^{2}-c^{2})(\beps_{t}^{2}-c^{2})-2E\beps_{t+d}^{2}\varepsilon_{t+1}(c+\beps_{t})\nonumber \\
 & =E(\beps_{t+d}^{2}-c^{2})(\beps_{t}^{2}-c^{2})-2cE\beps_{t+d}^{2}\varepsilon_{t+1}-2E\beps_{t+d}^{2}\varepsilon_{t+1}\beps_{t}\nonumber \\
 & =E\beps_{t+d}^{2}\beps_{t}^{2}-c^{4}-2cE\beps_{t+d}^{2}\varepsilon_{t+1}-2E\beps_{t+d}^{2}\varepsilon_{t+1}\beps_{t}.\label{eq: computation 3}
\end{align}

The rest of the proof proceeds in three steps.

\textbf{Step 1:} Compute $\gamma_{d}$ for $1\leq d\leq m-1$.

Define $\xi_{1}=\sum_{s=t-m+1}^{t-m+d}\varepsilon_{s}$, $\xi_{2}=\sum_{s=t-m+d+1}^{t}\varepsilon_{s}$
and $\xi_{3}=\sum_{s=t+1}^{t+d}\varepsilon_{s}$. These three quantities
are well defined because $1\leq d\leq m-1$. Notice that $\xi_{1}$,
$\xi_{2}$ and $\xi_{3}$ are mutually independent with mean zero
and satisfy $E\xi_{1}^{2}=d$, $E\xi_{2}^{2}=m-d$ and $E\xi_{3}^{2}=d$.
Moreover, $\beps_{t+d}=(\xi_{2}+\xi_{3})/m$ and $\beps_{t}=(\xi_{1}+\xi_{2})/m$.
Therefore, we have 
\begin{align*}
E\beps_{t+d}^{2}\beps_{t}^{2} & =m^{-4}E(\xi_{1}+\xi_{2})^{2}(\xi_{2}+\xi_{3})^{2}\\
 & =m^{-4}E(\xi_{1}^{2}+\xi_{2}^{2}+2\xi_{1}\xi_{2})(\xi_{2}^{2}+\xi_{3}^{2}+2\xi_{2}\xi_{3})\\
 & =m^{-4}\left(E\xi_{1}^{2}E\xi_{2}^{2}+E\xi_{1}^{2}E\xi_{3}^{2}+E\xi_{2}^{4}+E\xi_{2}^{2}E\xi_{3}^{2}\right)\\
 & =m^{-4}\left(2dm-d^{2}+E\xi_{2}^{4}\right).
\end{align*}

Notice that $E\xi_{2}^{4}=\sum_{s_{1}\neq s_{2}}E\varepsilon_{s_{1}}^{2}E\varepsilon_{s_{2}}^{2}+\sum_{s}E\varepsilon_{s}^{4}=(m-d)(m-d-1)+(m-d)\kappa_{2}$.
Thus, we have 
\begin{equation}
E\beps_{t+d}^{2}\beps_{t}^{2}=m^{-4}\left(2dm-d^{2}+(m-d)(m-d-1)+(m-d)\kappa_{2}\right)=m^{-4}\left(m^{2}+(m-d)(\kappa_{2}-1)\right).\label{eq: computation 4}
\end{equation}

Since $\beps_{t+d}=(\xi_{2}+\xi_{3})/m$, we have 
\begin{equation}
E\beps_{t+d}^{2}\varepsilon_{t+1}=m^{-2}E(\xi_{2}^{2}+\xi_{3}^{2}+2\xi_{2}\xi_{3})\varepsilon_{t+1}=m^{-2}E\xi_{3}^{2}\varepsilon_{t+1}=m^{-2}E\xi_{t+1}^{3}=m^{-2}\kappa_{1}.\label{eq: computation 5}
\end{equation}

We observe that 
\begin{align}
E\beps_{t+d}^{2}\varepsilon_{t+1}\beps_{t} & =m^{-3}E(\xi_{2}+\xi_{3})^{2}\varepsilon_{t+1}(\xi_{1}+\xi_{2})\nonumber \\
 & =m^{-3}E(\xi_{2}^{2}+\xi_{3}^{2}+2\xi_{2}\xi_{3})(\xi_{1}+\xi_{2})\varepsilon_{t+1}\nonumber \\
 & =m^{-3}E(\xi_{2}^{2}+\xi_{3}^{2}+2\xi_{2}\xi_{3})\xi_{2}\varepsilon_{t+1}\nonumber \\
 & =2m^{-3}E\xi_{2}^{2}\xi_{3}\varepsilon_{t+1}\nonumber \\
 & =2m^{-3}E\xi_{2}^{2}E\xi_{3}\varepsilon_{t+1}=2m^{-3}(m-d).\label{eq: computation 6}
\end{align}

Now we combine (\ref{eq: computation 3}) with (\ref{eq: computation 4}),
(\ref{eq: computation 5}) and (\ref{eq: computation 6}), obtaining
that for $1\leq d\leq m-1$, 
\begin{align}
\gamma_{d} & =m^{-4}(m-d)(\kappa_{2}-1)-2cm^{-2}\kappa_{1}-4m^{-3}(m-d)\nonumber \\
 & =\left[m^{-4}(\kappa_{2}-1)-4m^{-3}\right](m-d)-2cm^{-2}\kappa_{1}.\label{eq: computation 7}
\end{align}

\textbf{Step 2:} Compute $\gamma_{d}$ for $d=m$.

We notice that $\beps_{t+m}$ and $\beps_{t}$ are independent. This
means that $E\beps_{t+d}^{2}\beps_{t}^{2}=m^{-2}=c^{4}$. Moreover,
$E\beps_{t+d}^{2}\varepsilon_{t+1}=m^{-2}\kappa_{1}$. Finally, $E\beps_{t+d}^{2}\varepsilon_{t+1}\beps_{t}=E\beps_{t+d}^{2}\varepsilon_{t+1}E\beps_{t}=0$.
It follows by (\ref{eq: computation 3}) that 
\begin{equation}
\gamma_{m}=-2cm^{-2}\kappa_{1}.\label{eq: computation 8}
\end{equation}

\textbf{Step 3:} Compute $\gamma_{d}$ for $d=0$.

We observe that 
\begin{multline*}
\gamma_{0}=E(\Delta L_{t+1})^{2}=E\left(\beps_{t}^{2}-c^{2}-2\varepsilon_{t+1}(c+\beps_{t})\right)^{2}=E\left(\beps_{t}^{2}-c^{2}\right)^{2}+4E\varepsilon_{t+1}^{2}(c+\beps_{t})^{2}\\
=E\beps_{t}^{4}-c^{4}+4E(c+\beps_{t})^{2}=E\beps_{t}^{4}-c^{4}+4(c^{2}+m^{-1})=E\beps_{t}^{4}-m^{-2}+8m^{-1}.
\end{multline*}

By a similar argument as in the computation for $E\xi_{2}^{4}$, we
can show that $E\beps_{t}^{4}=m^{-3}(m-1+\kappa_{2})$. It follows
that 
\begin{equation}
\gamma_{0}=m^{-3}(\kappa_{2}-1)+8m^{-1}.\label{eq: copmutation 9}
\end{equation}

Now we apply (\ref{eq: computation 7}), (\ref{eq: computation 8})
and (\ref{eq: copmutation 9}) and compute the long-run variance
\begin{align*}
\Gamma_{\infty} & =\gamma_{0}+2\sum_{d=1}^{\infty}\gamma_{d}\\
 & =\gamma_{0}+2\sum_{d=1}^{m}\gamma_{d}\\
 & =m^{-3}(\kappa_{2}-1)+8m^{-1}\\
 & \qquad+2\sum_{d=1}^{m-1}\left\{ \left[m^{-4}(\kappa_{2}-1)-4m^{-3}\right](m-d)-2cm^{-2}\kappa_{1}\right\} +2\times(-2cm^{-2}\kappa_{1})\\
 & =m^{-3}(\kappa_{2}-1)+8m^{-1}-4cm^{-1}\kappa_{1}+2\left[m^{-4}(\kappa_{2}-1)-4m^{-3}\right]\sum_{d=1}^{m-1}(m-d)\\
 & =m^{-3}(\kappa_{2}-1)+8m^{-1}-4cm^{-1}\kappa_{1}+2\left[m^{-4}(\kappa_{2}-1)-4m^{-3}\right]\times\frac{m(m-1)}{2}\\
 & =4m^{-1}-4cm^{-1}\kappa_{1}+m^{-2}(\kappa_{2}+3).
\end{align*}

By the law of large numbers, $n^{-1}\sum_{t=m+1}^{m+n}(\Delta L_{t+1})^{2}$
converges in probability to $E(\Delta L_{t+1})^{2}=\gamma_{0}$. Therefore,
since $c=m^{-1/2},$ the test statistic would have an asymptotic variance
equal to 
\[
\frac{\Gamma_{\infty}}{\gamma_{0}}=\frac{4m^{-1}-4cm^{-1}\kappa_{1}+m^{-2}(\kappa_{2}+3)}{m^{-3}(\kappa_{2}-1)+8m^{-1}}=\frac{4m^{2}-4m^{3/2}\kappa_{1}+m(\kappa_{2}+3)}{8m^{2}+\kappa_{2}-1}.
\]

The proof is complete. 
\end{proof}
\begin{proof}[\textbf{Proof of Proposition \ref{prop: sub-t}}]
Let $Z_{k}=Q^{-1}\sum_{t=m+(k-1)Q+1}^{m+kQ}\Delta L_{t+1}$, where
$Q=n/K$. For simplicity, assume that $Q$ is an integer. By Theorem
5.20 of \citet{White2001} and the Cramer-Wold device, $\sqrt{Q}(Z_{1},...,Z_{K})$
converges in distribution to $(\xi_{1},...,\xi_{K})\sim N(0,c^{2}I_{K})$,
where $c^{2}=E(\Delta L_{t})^{2}+2\sum_{s=1}^{\infty}E(\Delta L_{t+s}\Delta L_{t})$.
Define the function $g$ by 
\[
g(\xi_{1},...,\xi_{K})=\frac{K^{1/2}\bar{\xi}}{\sqrt{(K-1)^{-1}\sum_{k=1}^{K}(\xi_{k}-\bar{\xi})^{2}}}
\]
with $\bar{\xi}=K^{-1}\sum_{k=1}^{K}\xi_{k}$. Then $S_{K}=g(\sqrt{Q}Z_{1},...,\sqrt{Q}Z_{K})$.
The desired result follows by the continuous mapping theorem. 
\end{proof}
\begin{proof}[\textbf{Proof of Proposition \ref{prop: example 1}}]
Define $B_{T}(r)=T^{-1/2}\sum_{s=1}^{\left\lfloor rT\right\rfloor }\varepsilon_{s}$.
Then by the functional central limit theorem (e.g., Theorem 7.13 of
\citet{White2001}), $B_{T}$ converges weakly to $B$, where $B(\cdot)$
is a standard Brownian motion. In fact, this weak convergence can
be strengthened to a strong approximation on a possible extended probability
space, i.e., $\sup_{x\in[0,1]}|B_{T}(x)-B(x)|=o_{P}(1)$; see e.g.,
Theorem 2.1.2 of \citet{csorgo2014strong}.

We notice that $\varepsilon_{t+1}=\sqrt{T}(B_{T}((t+1)/T)-B_{T}(t/T))$
and $f_{1,t}=\sqrt{T}t^{-1}B_{T}(t/T)$. Hence, 
\begin{align}
\Delta L_{t+1} & =(\varepsilon_{t+1}-\sqrt{T}t^{-1}B_{T}(t/T))^{2}-(\varepsilon_{t+1}-t^{-1/2})^{2}\nonumber \\
 & =Tt^{-2}B_{T}^{2}(t/T)-t^{-1}-2\varepsilon_{t+1}(\sqrt{T}t^{-1}B_{T}(t/T)-t^{-1/2})\nonumber \\
 & =Tt^{-2}B_{T}^{2}(t/T)-t^{-1}-2\sqrt{T}\left[B_{T}((t+1)/T)-B_{T}(t/T)\right]\left(\sqrt{T}t^{-1}B_{T}(t/T)-t^{-1/2}\right).\label{eq: rewriting Delta L}
\end{align}

Therefore, 
\begin{align}
\sum_{t=m}^{T-1}\Delta L_{t+1} & =\sum_{t=m}^{T-1}\left(Tt^{-2}B_{T}^{2}(t/T)-t^{-1}-2\sqrt{T}\left[B_{T}((t+1)/T)-B_{T}(t/T)\right]\left(\sqrt{T}t^{-1}B_{T}(t/T)-t^{-1/2}\right)\right)\nonumber \\
 & =\sum_{t=m}^{T-1}\left((t/T)^{-2}B_{T}^{2}(t/T)-(t/T)^{-1}\right)T^{-1}\nonumber \\
 & \qquad\qquad-2\sum_{t=m}^{T-1}\left[B_{T}((t+1)/T)-B_{T}(t/T)\right]\left((t/T)^{-1}B_{T}(t/T)-(t/T)^{-1/2}\right)\nonumber \\
 & =\int_{\lambda}^{1}\left(u^{-2}B^{2}(u)-u^{-1}\right)du-2\int_{\lambda}^{1}\left(u^{-1}B(u)-u^{-1/2}\right)dB(u)+o_{P}(1).\label{eq: numerator}
\end{align}

Similarly, 
\begin{align*}
 & \sum_{t=m}^{T-1}(\Delta L_{t+1})^{2}\\
 & =\underset{A_{1,T}}{\underbrace{\sum_{t=m}^{T-1}\left(Tt^{-2}B_{T}^{2}(t/T)-t^{-1}\right)^{2}}}+\underset{A_{2,T}}{\underbrace{4T\sum_{t=m}^{T-1}\left[B_{T}((t+1)/T)-B_{T}(t/T)\right]^{2}\left(\sqrt{T}t^{-1}B_{T}(t/T)-t^{-1/2}\right)^{2}}}\\
 & \qquad-\underset{A_{3,T}}{\underbrace{4\sqrt{T}\sum_{t=m}^{T-1}\left[B_{T}((t+1)/T)-B_{T}(t/T)\right]\left(\sqrt{T}t^{-1}B_{T}(t/T)-t^{-1/2}\right)\left(Tt^{-2}B_{T}^{2}(t/T)-t^{-1}\right)}}.
\end{align*}

Notice that 
\begin{align*}
A_{1,T} & =T^{-1}\sum_{t=m}^{T-1}\left((t/T)^{-2}B_{T}^{2}(t/T)-(t/T)^{-1}\right)^{2}T^{-1}\\
 & =T^{-1}\int_{\lambda}^{1}\left(u^{-2}B^{2}(u)-u^{-1}\right)^{2}du+o_{P}(1)=o_{P}(1)
\end{align*}
and 
\begin{align*}
A_{2,T} & =4\sum_{t=m}^{T-1}\left[B_{T}((t+1)/T)-B_{T}(t/T)\right]^{2}\left((t/T)^{-1}B_{T}(t/T)-(t/T)^{-1/2}\right)^{2}\\
 & =4\int_{\lambda}^{1}\left(u^{-1}B(u)-u^{-1/2}\right)^{2}du+o_{P}(1).
\end{align*}

Finally, we observe that 
\begin{align*}
 & A_{3,T}\\
 & =4\sum_{t=m}^{T-1}\left[B_{T}((t+1)/T)-B_{T}(t/T)\right]\left((t/T)^{-1}B_{T}(t/T)-(t/T)^{-1/2}\right)\left((t/T)^{-2}B_{T}^{2}(t/T)-(t/T)^{-1}\right)T^{-1}\\
 & =4\int_{\lambda}^{1}\left(u^{-1}B(u)-u^{-1/2}\right)\left(u^{-2}B^{2}(u)-u^{-1}\right)dB(u)du+o_{P}(1)=o_{P}(1).
\end{align*}

The above four displays imply that 
\[
\sum_{t=m}^{T-1}(\Delta L_{t+1})^{2}=4\int_{\lambda}^{1}\left(u^{-1}B(u)-u^{-1/2}\right)^{2}du+o_{P}(1).
\]

Therefore, the desired result follows by (\ref{eq: numerator}).
\end{proof}

\bibliographystyle{apalike}
\bibliography{reference_GW_test}

\end{document}